\documentclass[prl,aps]{revtex4}

\usepackage{latexsym}
\usepackage{epsfig}
\usepackage{amssymb}

\newcommand {\be}[1] {\begin{equation}\label{#1}}
\newcommand {\ee} {\end{equation}}
\newcommand {\ba}[1] {\begin{eqnarray}\label{#1}}
\newcommand {\ea} {\end{eqnarray}}
\newcommand {\eq}[1] {eq.~(\ref{#1})}
\newcommand {\fig}[1] {fig.~\ref{#1}}
\def \bp {\mbox{\boldmath $\partial$}}

\def \F2 {FPL${}^2$ }

\def \OMIT #1{}
\def \rem #1 {{\it #1}}

\begin{document}

\title {Continuous melting of compact polymers}
\author{Jesper Lykke Jacobsen}
\affiliation{LPTMS, Universit\'e Paris-Sud, B\^atiment 100, F-91405 Orsay, FRANCE}
\author{Jan\'{e} Kondev}
\affiliation{Physics Department, Brandeis University, Waltham, MA 02454, USA}

\date{\today}

\begin{abstract}

The competition between chain entropy and bending rigidity in compact polymers
can be addressed within a lattice model introduced by P.J.~Flory in 1956. It
exhibits a transition between an entropy dominated disordered phase and an
energetically favored crystalline phase. The nature of this order-disorder
transition has been debated ever since the introduction of the model. Here we
present exact results for the Flory model in two dimensions relevant for
polymers on surfaces, such as DNA adsorbed on a lipid bilayer. We predict a
continuous melting transition, and compute exact values of critical exponents
at the transition point.

\end{abstract}
\maketitle

Condensed phases of polymers are ubiquitous in the living world. Examples
range from the organization of DNA in viruses, to the native state of globular
proteins. For example, DNA condensation is required in order to fit the
complete viral genome, which can be tens of microns long, into the viral
capsid typically tens of nanometers in diameter. As the persistence length of
the DNA (50 nm) is comparable to the linear dimension of the viral capsid
(10--100 nm), one would expect it to pack into ordered, crystalline
configurations. Indeed, cryo-electron-microscopy studies of the T7 virus have
revealed a circular, inverse-spool arrangement of DNA in the capsid
\cite{cerritelli}. The observed condensed state of DNA is a result of many
competing effects such as chain and solvent entropy, electrostatic
interactions, bending stiffness of the polymer, and interaction with capsid
proteins \cite{riemer}.

%Similarly, globular proteins in their native state assume compact
%conformations which are necessary for their function. These examples raise
%important questions regarding possible condensed phases of polymers, and the
%nature of the transitions between them.

%Although realistic models of proteins are generally intractable analytically,
%it is nevertheless of interest to obtain exact results for models in which a
%subset of all these interactions compete.

%Motivated by these observations, we here study the
%competition between chain entropy and bending energy of a polymer chain under
%extreme confinement, when only compact configurations are allowed. To study
%the transition between an entropy dominated disordered state and an
%energetically favored crystalline one,  we turn to a
%simple lattice model of a compact semiflexible polymer. The model we employ
%was first introduced by Flory almost 50 years ago in the context of polymer
%melting \cite{flory}. It has also served as the starting point for
%constructing simple lattice models of protein folding \cite{dill}.

These observations motivate the study of the
competition between chain entropy and bending energy of a polymer chain under
extreme confinement, when only compact configurations are allowed. The
resulting transition between an entropy dominated disordered state and an
energetically favored crystalline one is captured by a
simple lattice model of a compact semiflexible polymer,
first introduced by Flory almost 50 years ago in the context of polymer
melting \cite{flory}. It has also served as the starting point for
constructing simple lattice models of protein folding \cite{dill}.

In this Letter, we provide an analytic solution of the Flory model in two
dimensions using field theoretical methods.
%(Note that two-dimensional
%polymers can be realized experimentally by adsorbing DNA on a lipid bilayer
%\cite{DNA}; whether the effect of compactness can be measured is an
%interesting open question.)
We resolve the long-standing debate
\cite{nagle,gujrati,yoon,orland,saleur,mansfield} about the nature of the
transition, where some numerical results \cite{yoon} and approximate analytic
approaches \cite{orland} were indicative of a first-order transition, while
other numerical work \cite{saleur,mansfield} and theoretical arguments based
on similarities with the exactly solvable six-vertex model
\cite{nagle,saleur}, seemed to imply a continuous transition. We show that the
melting transition of a single semiflexible compact polymer is continuous. Furthermore,
we characterize the transition precisely by calculating the exact values of
the critical exponents that accompany it.

\begin{figure}
\begin{center}
 \leavevmode
 \epsfysize=86mm{\epsffile{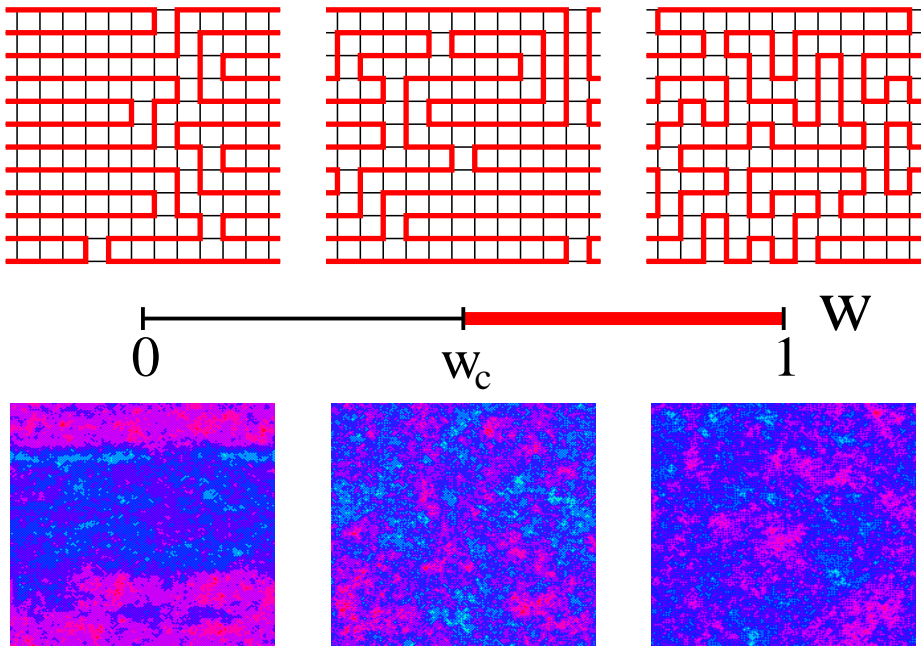}}
 \end{center}
 \protect\caption[3]{Phase diagram of the Flory model. The Boltzmann weight of a single
 bend in the chain is $w$. For $w=1$ the model describes
 a flexible compact polymer with many bends present in a typical polymer configuration.
 For $w<1$ there is an energy penalty associated with bending and we are dealing with a
 semiflexible chain. For $w_{\rm c}\le w\le 1$ the chain is critical. At $w=w_{\rm c}$ there is a
 continuous transition to an ordered chain configuration; the
 order parameter is the difference between the number of vertical and horizontal links.
 The polymer configurations shown in the top three panels (with periodic boundary conditions imposed
 in the horizontal direction) are typical for $w=1/3$, $w=w_{\rm c}\approx1/2$, and $w=1$.
 The density plots below them are of the second height component in
 the related loop model, for the same values of $w$, clearly showing the transition from a rough
 to a flat interface. }
\label{fig:phasediagram}
\end{figure}

\paragraph{Semiflexible loop model}
In the Flory model a configuration of a polymer chain is described by a random walk on a
square lattice (cubic, in three dimensions) which visits every site of the lattice exactly once. The energy
associated with a configuration is $B \epsilon>0$, where $B$ is the number of  bends in the chain.
At large temperatures ($k_B T\gg \epsilon$) bends proliferate for entropic reasons, thus reducing
the persistence length of the polymer. In the
low-temperature limit ($k_B T\ll \epsilon$)
bends are very unlikely and the persistence
length is large, of the order of the lattice size, as seen in \fig{fig:phasediagram}.
The central question we address here is the nature of the transition
between the disordered high-temperature phase and the ordered low-temperature phase.

The interplay between the entropy and the energy is encoded in the partition function of the
Flory model
\be{eq:partition}
Z_F = \sum_{\Gamma_F} w^{B(\Gamma_F)} \ ,
\ee
where $w=\exp(-\epsilon/k_B T)$ is the Boltzmann weight associated with a bend, and $B(\Gamma_F)$ is
the number of 90${}^\circ$ bends in the compact chain configuration $\Gamma_F$.
For $T\to 0$ ($w\to 0$) the equilibrium state is one that minimizes $B$, while for $T\to\infty$ ($w\to 1$)
entropy dominates and there is a large, macroscopic number of bends in a typical chain configuration;
see \fig{fig:phasediagram}.
In order to investigate the transition between the high and low temperature
phase we construct a mapping of the polymer problem to a lattice model of fluctuating loops,
following Nienhuis \cite{nienrev}.

We define the semiflexible loop model (SFL), whose configurations $\Gamma_L$ are all possible ways of covering the
square lattice with non-intersecting loops. The partition function of the SFL model reads
\be{partition}
Z_L = \sum_{\Gamma_L} n^{N(\Gamma_L)} w^{B(\Gamma_L)}  \ ,
\ee
where each of the $N(\Gamma_L)$ loops in the configuration $\Gamma_L$ carries
a weight $n>0$, and bends are weighted by $w$ as in the Flory model. Note that
the $n\to 0$ limit of $Z_L/n$ yields the Flory model partition
function for a single {\em closed} compact polymer.
%(Later we will show how the solution for the closed polymer problem provides
%a solution for the open chain case as well.)
Instead of explicitly computing $Z_L$
the approach we take is to construct a continuum limit of the SFL model in the form of a
field theory. Computations of correlation functions in this field theory lead to the
phase diagram and exact scaling exponents for any $n$, including the polymer ($n\to 0$) limit.

\paragraph{Height model}
Since loops in $\Gamma_L$ do not intersect they can be interpreted as contour lines of the height of an interface.
In the SFL model, the height is a
three component scalar field. To see this, we first note that bonds not covered
by loops form so-called {\em ghost} loops, if periodic boundary conditions are imposed. Then we assign to each
real and ghost loop an orientation (clockwise or anticlockwise) independently and randomly.
By doing so, each bond of the square lattice is found in one of four states,
labeled ${\bf A}, {\bf B}, {\bf C},$ or ${\bf D}$, depending on whether it is covered  by a real or a ghost loop
oriented in one or the other direction. We parametrize $n=2\cos(\pi e)$,
with $0 \le e < 1/2$, and assign a weight $\exp(\pm i \pi e)$ (resp.~$\exp(\pm i \pi/3)$) to each orientation of a real (resp.~ghost) loop.
Tracing over the two loop orientations, we recover the desired weights:
$n$ for the real loops and $1$ for the ghost loops.

The heights for the SFL model  are defined at the centers of the elementary square plaquettes of the lattice.
Height differences between neighboring plaquettes are given by {\em vectors} ${\bf A}, {\bf B}, {\bf C},$ or ${\bf D}$.
Since each vertex is shared by four bonds, all necessarily in different states,
a single relation,
${\bf A}+{\bf B}+{\bf C}+{\bf D}=0$, results from requiring that the total height increment when
traversing a closed lattice path be zero.
Thus, we conclude that
out of the four vectors three are linearly independent and that the heights are three dimensional.
For the remainder of the paper we adopt the following normalization:
${\bf A} = (-1,+1,+1)$, ${\bf B} = (+1,+1,-1)$, ${\bf C} = (-1,-1,-1)$, and ${\bf D} = (+1,-1,+1)$.
The partition function of the SFL model can now be rewritten as a sum over height configurations,
since a height configuration uniquely specifies an oriented loop configuration.

The continuum limit of the SFL model is obtained by coarse-graining the height field over
domains much smaller than the system size and much bigger than a single plaquette. This  yields a continuous
height function ${\bf h}({\bf x})=\left( h^1({\bf x}), h^2({\bf x}), h^3({\bf x})\right)$
defined over the plane ${\bf x}$. For $n\le 2$ the height field is assumed rough \cite{JK_npb} and the
partition function  can be written as a sum (path integral)
over all height functions with weights given by $\exp(-S)$, where
\be{action_el2}
S = \frac{1}{2} \int \! {\rm d}^2{\bf x} \  K_{ij}(\bp h^i \cdot \bp h^j) +
    \frac{{\rm i}}{4 \pi} \int \! {\rm d}^2{\bf x} \ ({\bf e}_0 \cdot {\bf h}) \rho({\bf x}) +
    \gamma \int \! {\rm d}^2{\bf x} \ \cos\left(2\pi h^2\right) \ .
\ee
This particular form of the coarse-grained free energy $S$ is dictated by the symmetries of the
loop model which involve translations in the space of heights, and permutations of the
height-difference vectors \cite{JK_npb}.
The value of the background charge
\be{bc}
 {\bf e}_0 = - \frac{\pi}{2} (1/3 + e, 0, 1/3 - e) \
\ee
follows from imposing the correct weighting of loops that close at infinity
($\rho({\bf x})$ is the curvature of the ${\bf x}$-surface), while the non-zero elastic
constants
\be{el_cst}
K_{11} = \frac{\pi}{8} (5/3-e), \ \
K_{13} = \frac{\pi}{8} (e-1/3), \  \
K_{22} = K_{22}(e,w)
\ee
are partially fixed by the loop ansatz which states that loop weights do not flow under renormalization \cite{jk_prl}.

In the previously studied infinite-temperature limit ($w\to 1$),
the elastic constant $K_{22}=\pi(1-e)/(5-3e)$ is also fixed by the loop ansatz,
and the cosine term is absent from $S$ \cite{JK_npb}. For $w<1$ the cosine potential accounts for the
energy penalty associated with bends. Its scaling dimension in the $w=1$ theory, $x_B=\pi/K_{22}=(5-3e)/(1-e)$,
is greater than two for  $0 \le e < 1/2$ ($0< n \le 2$), and is therefore an {\em irrelevant} perturbation. It has the effect of
renormalizing the elastic constant $K_{22}$ which ultimately drives the transition to the crystalline
phase at $w=w_c$.

\paragraph{Phase diagram and critical exponents}
The field theory of the SFL model implies the following phase diagram for any $n\le 2$, including
the polymer limit $n\to 0$. For $w=1$ the interface described by the height field is {\em rough} and
the loop model is critical, i.e., it is characterized by a power-law distribution of loop sizes. As $w$ is reduced, the bending stiffness of the loops increases,
and the height remains rough until a critical value $w=w_{\rm c}$. At $w_{\rm c}$  there is a {\em roughening transition} \cite{weeks}
to a smooth interface; see \fig{fig:phasediagram}. In the loop model this corresponds to a phase with a
vanishing density of bends. The roughening transition is
continuous, in the Kosterlitz-Thouless universality class. At the transition the cosine term in
$S$ becomes relevant in the renormalization group sense \cite{weeks}. In other words, the scaling dimension of the
operator $\cos(2\pi h^2)$ at $w=w_{\rm c}$ is 2. Computing this dimension using the field theory defined by $S$
then leads to the prediction $K_{22}(e,w_{\rm c}) = \pi/2$. One important test of this result comes from the
$n=1$ ($e=1/3$) SFL model, which maps to the exactly solvable six-vertex model. In this case $w_{\rm c}=1/2$ and
$K_{22}= \arcsin(1/2w)$ have been computed previously \cite{baxter}.

In the polymer limit the transition described by the field theory takes us from a critical geometry
of a disordered compact polymer, at low bending stiffness, to an ordered state, at high bending stiffness;
see \fig{fig:phasediagram}. The critical phase is characterized by scaling exponents $\nu$ and
$\theta$. Namely, the probability distribution $p(r,l)$, for the end-to-end distance $r$ of a polymer of length $l$,
%and end-to-end distance $r$
%Namely, in the limit that $r$ is much greater than the lattice spacing,
%but much less than the size of the box the polymer is confined in, the probability
has a scaling form $p(r,l) = r^{\theta} f(r/l^\nu)$ \cite{degennes}.
Using the field theory defined by \eq{action_el2} these exponents can be computed exactly \cite{JK_npb}.
We find
\be{exponents}
\nu = \frac{1}{2}, \ \ {\rm and } \  \ \theta= \frac{3\pi - 16 K_{22}}{8 \pi} \ ;
\ee
note that $\theta$ varies continuously with
$w$ via the elastic constant $K_{22}$. The calculation of $\nu$ gives another important
check on the field theory since $\nu=1/2$ follows directly from the fact that compact polymers completely
fill the plane. Finally, exact values for $K_{22}$
in the $w=1$ and $w=w_{\rm c}$ case, lead to predictions $\theta(w=1)= 5/56$ and $\theta(w=w_{\rm c})=-5/8$.
These values of $\theta$
have a simple physical interpretation. Namely, in the absence of bending stiffness ($w=1$) the ends of the
chain repel (positive $\theta$), as is the case for open chain conformations of polymers in a good solvent.
Somewhat surprisingly, this entropic repulsion decreases and turns to attraction (negative $\theta$) as the
chain stiffness is increased.

\paragraph{Numerical transfer matrix results}
The main result of this paper, the prediction of continuous melting of compact polymers in two-dimensions,
rests on the validity of the proposed field theory of the Flory model, \eq{action_el2}.
To further test the field theory we have made use of a numerical transfer matrix approach \cite{JK_npb} to compute
the elastic constant $K_{22}$ and the exponent $\theta$. Results are shown in table~\ref{tab}. We
see that \eq{exponents} is satisfied to a very good approximation and that the exact values for the
two quantities at $w=1$ and $w=w_{\rm c}$ are confirmed by the numerical computations.

\begin{table}
 \begin{center}
\begin{tabular}{|c||c|l|l|} \hline
  $w^{-1}$ & $K_{22}$ & \ $\theta_{\rm th}$  & $\theta_{\rm num}$ \\ \hline
  $1$ & $\pi/7$  & \ \ $5/56$ & \ \ $0.088$ \\
  $1.2$ & $0.584$  & \ \ $0.004$  & \ \ $0.004$ \\
  $1.4$ & $0.736$   & $-0.094$ & $-0.096$ \\
  $1.6$ & $0.924$ & $-0.214$  & $-0.22$  \\
  $1.8$ & $1.210$  & $-0.404$ & $-0.40$ \\
  $w_{\rm c}^{-1}\approx 2$ & $\pi/2$ & $-5/8$ & $-0.7$ \\ \hline
\end{tabular}
\end{center}
 \protect\caption[2]{\label{tab} Numerical transfer matrix results. Exact (for $w=1,w_{\rm c}$) and numerically
 determined values of the elastic constant
 $K_{22}$ were used to compute the exponent $\theta_{\rm th}$ from \eq{exponents}, for various values of the bend weight
 $w$; this compares favorably with the direct numerical determination $\theta_{\rm num}$.}
\end{table}

\paragraph{Experiments}
DNA absorbed on a lipid bilayer provides a laboratory for testing theories of two-dimensional polymers.
Previous experiments \cite{DNA} on this system have measured the theoretically predicted scaling exponents
derived from the two-dimensional self-avoiding walk model. These measurements were done at
low DNA concentrations. As the amount of DNA on the surface is increased it was observed that
single chain conformations become more compact. To study the transition discussed here
compaction to areas of linear dimensions approaching the DNA persistence length would be needed. It
would then be interesting to monitor the two ends of the DNA using fluorescent labels, as the
amount of compaction is increased. This would provide direct information about $p(r,l)$ and the
exponent $\theta$ computed above.

\paragraph{Acknowledgements} We are grateful to K.A.~Dill for introducing us to the Flory model.
JK further thanks the KITP in Santa Barbara, where this work was initiated, for hospitality.
JK is supported by the NSF under grant number DMR-9984471 and is a Cottrell Scholar of Research Corporation.


\begin{thebibliography}{10}

\bibitem{cerritelli} M.E.\ Cerritelli {\em et al.},
    {\em Cell}  {\bf 91}, 271-280 (1997).

\bibitem{riemer} S.C.\ Riemer and V.A.\ Bloomfield,
    {\em Biopolymers} {\bf 17}, 785-794 (1978).

\bibitem{flory} P.J.\ Flory,
    {\em Proc. R. Soc. London, Ser. A} {\bf 234}, 60-73 (1956).

\bibitem{dill} H.S.\ Chan and K.A.\ Dill,
    {\em Macromolecules} {\bf 22}, 4559-4573 (1989).

\bibitem{nagle} J.F.\ Nagle,
    {\em Proc. R. Soc. London, Ser. A} {\bf 337}, 569-589 (1974).

\bibitem{gujrati} P.D.\ Gujrati and M.N.\ Goldstein,
    {\em J. Chem. Phys.} {\bf 74}, 2596-2603 (1981).

\bibitem{yoon} A.\ Baumgartner and D.\ Yoon,
    {\em J. Chem. Phys.} {\bf 79}, 521-522 (1983).

\bibitem{orland} J.\ Bascle, T.\ Garel, H.\ and Orland,
    {\em J.~Phys.~A:~Math. Gen.} {\bf 25}, L1323-L1329 (1992).

\bibitem{saleur} H.\ Saleur,
    {\em J. Phys. A: Math. Gen.} {\bf 19}, 2409-2423 (1986).

\bibitem{mansfield} M.L.\ Mansfield,
    {\em Macromolecules} {\bf 27}, 4699-4704 (1994).

\bibitem{nienrev} B.~Nienhuis, in {\em Phase Transitions and Critical
                          Phenomena}, edited by C.~Domb and J.~L.~Lebowitz
                          (Academic, London, 1987), Vol.~11.

\bibitem{JK_npb} J.L.\ Jacobsen and J.\ Kondev,
    {\em Nucl.~Phys.~B} {\bf 532}, 635-688 (1998).

\bibitem{jk_prl} J.\ Kondev,
    {\em Phys. Rev. Lett.} {\bf 78}, 4320-4323 (1997).

\bibitem{weeks} S.T.\ Chui and J.D.\ Weeks,
    {\em Phys. Rev. B} {\bf 14}, 4978-4982 (1976);
    J.V.\ Jos\'e, L.P.\ Kadanoff, S.\ Kirkpatrick, D.R.\ Nelson,
    {\em Phys. Rev. B} {\bf 16}, 1217-1241 (1977).

\bibitem{baxter} R.J.\ Baxter,
 {\em Exactly Solved Models in Statistical Mechanics}
                         (Academic Press, New York, 1982).

\bibitem{degennes} P.-G.\ de Gennes, {\em Scaling Concepts in Polymer
                          Physics} (Cornell University Press, Ithaca, 1979).

\bibitem{DNA} B.\ Miller and J.O.\ R\"adler,
    {\em Phys. Rev. Lett.} {\bf 82}, 1911-1914 (1999).

\end{thebibliography}
\end{document}